\documentclass[aps,prd,twocolumn,superscriptaddress,groupedaddress,nofootinbib]{revtex4}

\usepackage[hidelinks=true]{hyperref}
\hypersetup{colorlinks=true,linkcolor=blue,citecolor=blue,urlcolor=blue}
\usepackage{amsmath,amssymb,bm,color,dcolumn,mathrsfs,tabularx,graphicx,Macro}

\begin{document}

\title{Testing parity-violating physics from cosmic rotation power reconstruction}

\author{Toshiya Namikawa}
\affiliation{Department of Physics, Stanford University, Stanford, California 94305, USA}
\affiliation{Kavli Institute for Particle Astrophysics and Cosmology, SLAC National Accelerator Laboratory, 2575 Sand Hill Road, Menlo Park, California 94025, USA}

\begin{abstract}
We study the reconstruction of the cosmic rotation power spectrum produced by parity-violating physics, with an eye to ongoing and near future cosmic microwave background (CMB) experiments such as BICEP Array, CMBS4, LiteBIRD and Simons Observatory. In addition to the inflationary gravitational waves and gravitational lensing, measurements of other various effects on CMB polarization open new window into the early Universe. One of these is anisotropies of the cosmic polarization rotation which probes the Chern-Simons term generally predicted by string theory. The anisotropies of the cosmic rotation are also generated by the primordial magnetism and in the Standard Model extension framework. The cosmic rotation anisotropies can be reconstructed as quadratic in CMB anisotropies. However, the power of the reconstructed cosmic rotation is a CMB four-point correlation and is not directly related to the cosmic-rotation power spectrum. Understanding all contributions in the four-point correlation is required to extract the cosmic rotation signal. Assuming inflationary motivated cosmic-rotation models, we employ simulation to quantify each contribution to the four-point correlation and find that (1) a secondary contraction of the trispectrum increases the total signal-to-noise, (2) a bias from the lensing-induced trispectrum is significant compared to the statistical errors in, e.g., LiteBIRD and CMBS4-like experiments, (3) the use of a realization-dependent estimator decreases the statistical errors by $10$\%--$20$\%, depending on experimental specifications, and (4) other higher-order contributions are negligible at least for near future experiments. 
\end{abstract}

\maketitle


\section{Introduction}

One of the most important goals in cosmic microwave background (CMB) polarization cosmology is to detect the $B$-mode polarization from the inflationary gravitational waves (GWs). Multiple CMB experiments have been searching for evidence of the inflationary GWs. The current best upper bound on the amplitude of the inflationary GWs is obtained as $r<0.07$ ($2\sigma$) by BICEP2 / Keck Array (BK) \cite{BKVI}. Still there is no evidence of the inflationary GWs.

In addition to the inflationary GWs, theories in the early Universe can be tested through various observational effects on CMB polarization. Measurements of the polarization rotation angle (cosmic rotation) are known to be a unique probe of new physics in the early Universe containing a pseudoscalar Nambu-Goldstone boson coupled with photons by the Chern-Simons electro-magnetic term (e.g., \cite{Carroll:1998,Pospelov:2009,Finelli:2009,Shiraishi:2016}). Such theory is known to be a generic prediction of string theory and detection of the pseudoscalar fields provides implications for fundamental physics (for review, see e.g. \cite{Marsh:2016} and references therein). The presence of the pseudoscalar fields leads to the cosmic birefringence and the CMB polarization angle, $\a$, is rotated. Multiple studies have discussed the fluctuations of the pseudo-scalar fields which produce spatial variation in $\a$ (e.g., \cite{Kamionkowski:2010,Caldwell:2011,Gluscevic:2012qv}). 

The cosmic-rotation measurement can be also used to probe the effect of the Faraday rotation by the primordial magnetic fields (PMFs) (e.g., \cite{Kosowsky:1996,Harari:1997}). While PMFs are also constrained by the fact that they produce vector and tensor perturbations and generate $B$-mode spectrum \cite{Shaw:2010}, the cosmic rotation induced by the Faraday rotation is sensitive to the flat spectrum (lower value of the spectral index) \cite{Yadav:2012b} and its direct measurement provides a complementary test of PMFs \cite{Pogosian:2014}. 

Ref.~\cite{Leon:2017} shows that the cosmic rotation is generated by the Standard Model extension  which has a parity-violating term. 

In CMB observations, the cosmic rotation of the CMB Q and U maps are described as (e.g., \cite{Gluscevic:2009})
\al{
	[\rQ\pm\iu\rU](\hatn) = \E^{\pm 2\iu \a(\hatn)}[Q\pm\iu U](\hatn) \,. \label{Eq:rotmap}
}
The cosmic rotation produces nonzero correlations between temperature and $B$ mode, and also between $E$ and $B$ modes \cite{Lue:1999}. Anisotropies of the cosmic rotation are induced by inhomogeneous pseudoscalar fields and create mode mixing between $E$ and $B$ modes which is similar to the gravitational lensing. The cosmic rotation anisotropies can be reconstructed from CMB maps through the mixing between different Fourier modes \cite{Kamionkowski:2009}. 

The cosmic rotation has been measured by multiple CMB experiments. The uniform cosmic rotation has been constrained as, e.g., $\a=-0.36\pm1.24({\rm stat.})\pm1.5({\rm sys.})$ [deg] by WMAP \cite{Hinshaw:2013}, and $\a=0.35\pm0.05({\rm stat.})\pm0.28({\rm sys.})$ [deg] recently by Planck \cite{P16:rot}. The anisotropic cosmic rotation has also been constrained by multiple studies (e.g., \cite{Kamionkowski:2010,Gluscevic:2012qv,Liu:2016dcg}). The current tightest upper bound on the amplitude of the scale-invariant rotation spectrum is $L(L+1)C^{\a\a}_L/(2\pi)\alt 1$deg$^2$ ($2\sigma$) by POLARBEAR collaboration (2015) \cite{PB15:rot} (here $C_L^{\a\a}$ and $L$ denote the cosmic rotation spectrum and multipole, respectively). There are still no evidence of the cosmic rotation, and future CMB polarization experiments such as BICEP Array \cite{BICEPArray}, CMBS4 \footnote{\url{https://cmb-s4.org/CMB-S4workshops/index.php/Main_Page}}, Simons Observatory \footnote{\url{https://simonsobservatory.org/}} and LiteBIRD \cite{LiteBIRD} will probe the cosmic rotation by significantly improving their sensitivity to CMB polarization. In these future experiments, the measurement of the uniform rotation is expected to suffer from the significant systematics uncertainties, and the anisotropies of the cosmic rotation would be an alternative probe to explore the parity-violating physics and primordial magnetism from CMB observations. 

In this paper, we discuss the reconstruction of the cosmic-rotation power spectrum and specify non-negligible contributions to the reconstructed power spectrum with a numerical simulation. While the method to reconstruct the cosmic-rotation fluctuations has been discussed with multiple works (e.g., \cite{Kamionkowski:2009,Yadav:2009}), the reconstruction of the cosmic-rotation power spectrum is not well explored. Similar to the gravitational lensing, the power spectrum of the estimator, $\esta$, is not equivalent to the cosmic-rotation power spectrum, $C_L^{\a\a}$, i.e. the power of $\esta$ is the four-point correlation of the observed CMB. The most significant contribution comes from the disconnected part of the four-point correlation (hereafter, disconnected bias) which is included in the above data analysis \cite{Gluscevic:2012qv,PB15:rot}. There are, however, further contributions to the reconstructed power spectrum which must be taken into account in ongoing and near future high-sensitivity polarization experiments. For example, a secondary contraction of the trispectrum in the presence of the cosmic rotation produces a nontrivial contribution (hereafter, N1 term). The N1 term is considered as a signal of the cosmic rotation and the expected signal-to-noise could be enhanced compared to ignoring the N1 term. The lensing-induced trispectrum could also contribute as a bias (hereafter, lensing bias). The study of quantifying these contributions is required to estimate the cosmic-rotation power spectrum in high-sensitivity polarization experiments. 

This paper is organized as follows. In \sec{sec:rec} we summarize the cosmic rotation anisotropies. In \sec{sec:pow} we study the reconstruction of the cosmic rotation power spectrum. \sec{sec:sum} is devoted to summary and conclusion.

Throughout this paper, we assume a flat $\Lambda$CDM model with the best-fit parameters from Planck \cite{P15:main}. 

\section{Cosmic rotation anisotropies} \label{sec:rec}

\subsection{Anisotropies of the cosmic rotation from parity-violating physics and primordial magnetism}

String theory generally predicts the presence of a pseudoscalar Nambu-Goldstone boson coupled to the Chern-Simons electromagnetic term described as (e.g., \cite{Marsh:2016})
\al{
	L \supset \frac{a}{2f_a}F_{\mu\nu}\widetilde{F}^{\mu\nu} \,.
}
Here $a$ is the pseudoscalar fields, $f_a$ is the coupling constant, $F_{\mu\nu}$ is the field strength of the electromagnetic fields, $\tilde{F}^{\mu\nu}$ is the dual of $F^{\mu\nu}$. The presence of the pseudoscalar fields leads to the cosmic birefringence and the CMB polarization angle is rotated as $\a = \Delta a/f_a$ where $\Delta a$ is the change of the pseudoscalar fields along photon's trajectory (e.g., \cite{Pospelov:2009}). If $a$ is effectively massless during inflation, the power spectrum of $\a$ in the large-scale limit is given as \cite{Caldwell:2011}
\al{
	\frac{L(L+1)}{2\pi}C_L^{\a\a} = \left(\frac{H_I}{2\pi f_a}\right)^2 \,. \label{Eq:claa}
}
Here $H_I$ is the Hubble expansion rate in the inflationary era. 

Another candidate of the origin of the cosmic rotation is the PMFs. The cosmic rotation induced by the PMFs is given as \cite{Kosowsky:1996,Harari:1997} 
\al{
	\a = \frac{3}{16\pi^2 e\nu^2}\Int{}{\bm{l}}{} \cdot \dot{\tau}\bm{b} \,,
}
where $\nu$ is the observed frequency, $e$ is the elementary charge, $\dot{\tau}$ is the differential optical depth, $\bm{b}$ is the comoving magnetic field strength, and $d\bm{l}$ is the comoving length element along the trajectory of CMB photons. Although the cosmic-rotation power spectrum depends on the model of the PMFs, the nearly scale-invariant spectrum of the PMFs generated at the inflation leads to approximately the scale-invariant form of \eq{Eq:claa} \cite{Pogosian:2011}. The relation between the effective magnetic field strength and the rotation power spectrum is then given as \cite{De:2013}
\al{
	\frac{L(L+1)}{2\pi}C_L^{\a\a} 
		= 2.3\times 10^{-5}\left(\frac{30 {\rm GHz}}{\nu}\right)^4
		\left(\frac{B}{1\,{\rm nG}}\right)^2 \,.
}

Since the cosmic rotation induced by the above two scenarios has the scale-invariant spectrum, this paper focus on the rotation power spectrum whose shape is described as 
\al{
	\frac{L(L+1)}{2\pi}C_L^{\a\a} = A_{CB}\times 10^{-5} \,. \label{Eq:claa:n=0}
}
Here, we introduce an amplitude parameter of the cosmic rotation power spectrum, $A_{CB}$. $A_{CB}<31$ ($95\%$ C.L.) is the current upper bound from Ref.~\cite{PB15:rot}. $A_{CB}\sim \mC{O}(1)$ can be tested by the BK experiment with data up to 2014 and we set $A_{CB}=1$ in our analysis.  

\subsection{Mode coupling induced by anisotropic cosmic rotation}

The reconstruction of the anisotropic cosmic rotation from CMB maps is based on the fact that anisotropies of the cosmic rotation produces off-diagonal mode coupling between the $E$ and $B$ modes. An estimator of the cosmic rotation anisotropies, $\a(\hatn)$, is given as a quadratic in CMB \cite{Kamionkowski:2009}. Here we summarize the method for reconstructing $\a(\hatn)$ from the CMB polarization. Hereafter, we use $L$ for the multipoles of the $\a$ and $\l$ for the $E$ and $B$ modes. 

The $E$ and $B$ modes are defined using the Stokes $Q$ and $U$ maps as
\al{
	E_{\bl} \pm i B_{\bl} = - \FT{\hatn}{\bl}{f} [Q\pm\iu U](\hatn) \E^{\mp 2\iu\varphi_{\bl}}  \,.
}
Here we denote $\varphi_{\bl}$ as the angle of the multipole vector, $\bl$, measured from the Stokes $Q$ axis. From \eq{Eq:rotmap}, the rotated $E$ and $B$ modes are given by (e.g., \cite{Kamionkowski:2009})
\al{
	\rE_{\bl} &= \tE_{\bl} + \Int{2}{\bL}{(2\pi)^2} 2\a_{\bL} 
		\notag \\
		&\times [ -\tE_{\bl-\bL}\sin 2\varphi_{\bl-\bL,\bl} - \tB_{\bl-\bL}\cos 2\varphi_{\bl-\bL,\bl} ]
		\label{Eq:rotated-E} \\
	\rB_{\bl} &= \tB_{\bl} + \Int{2}{\bL}{(2\pi)^2} 2\a_{\bL}
		\notag \\
		&\times [ \tE_{\bl-\bL}\cos 2\varphi_{\bl-\bL,\bl} - \tB_{\bl-\bL}\sin 2\varphi_{\bl-\bL,\bl} ]
	\,, \label{Eq:rotated-B}
}
where $\varphi_{\bl_1,\bl_2}\equiv\varphi_{\bl_1}-\varphi_{\bl_2}$. In the above, we also assume that the higher order of $\a$ is ignored, i.e., the rotated CMB map of \eq{Eq:rotmap} is given by 
\al{
	[\rQ\pm\iu\rU](\hatn) \simeq [1\pm 2\iu \a(\hatn)][Q\pm\iu U](\hatn) 
	\,. \label{Eq:linrot}
}
The rotation-induced mode coupling between $E$ and $B$ modes are then given as \cite{Yadav:2009}
\al{
	\ave{\rE_{\bL}\rB_{\bl-\bL}}_{\rm CMB} = w_{\bl,\bL}^\a \a_{\bl} 
	\,. \label{Eq:weight}
}
Here the ensemble average, $\ave{\cdots}\rom{CMB}$, operates on unrotated CMB anisotropies under a fixed realization of the cosmic rotation fields, $\a$. The weight functions is then given by
\al{
	w^\a_{\bL,\bl} = 2(\tCEE_\l-\tCBB_{|\bL-\bl|})\cos 2\varphi_{\bl,\bL-\bl} 
	\,, \label{Eq:weight:a}
}
where $\tCEE_\l$ and $\tCBB_\l$ are the lensed $E$- and $B$-mode power spectrum, respectively. In this paper, we only consider the mode coupling between $E$ and $B$ modes since the $EB$ estimator is the best to constrain the cosmic rotation \cite{Yadav:2012a}. 

\subsection{Quadratic estimator of anisotropic cosmic rotation}

\eq{Eq:weight} motivates the following estimator for the rotation angle \citep{Yadav:2009} 
\al{
	\esta_{\bL} = A_{\bL}^\a (\uesta_{\bL}-\ave{\uesta_{\bL}})  \,, \label{Eq:est}
}
where $\ave{\cdots}$ is the ensemble average over realizations of observed $E$ and $B$ modes and $\uesta_{\bL}$ is the unnormalized $EB$ estimator,
\al{
	\uesta_{\bL} &= \Int{2}{\bl}{(2\pi)^2} w^\a_{\bL,\bl}
		\frac{\hE_{\bl}}{\hCEE_\l}\frac{\hB_{\bL-\bl}}{\hCBB_{|\bL-\bl|}} 
	\,. \label{Eq:uest}
}
Here $w^\a_{\bL,\bl}$ is the weight function given at \eq{Eq:weight:a}. The second term, $\ave{\uesta_{\bL}}$, is a correction for the mean-field bias and is usually estimated from the simulations. The quantities, $\hE$ and $\hB$, are the observed $E$ and $B$ modes which contain the instrumental noise. Their power spectra are denoted as $\hCEE_\l$ and $\hCBB_\l$, respectively. The quantity $A_{\bL}$ is the quadratic estimator normalization and is given as
\al{
	A^\a_{\bL} = \left[\Int{2}{\bl}{(2\pi)^2} 
		\frac{w_{\bL,\bl}^2}{\hCEE_{\l}\hCBB_{|\bL-\bl|}}\right]^{-1}
	\,. \label{Eq:norm}
}
The weight function of $\a$ is orthogonal to that of the lensing potential and the estimator of \eq{Eq:est} is not biased by the presence of the lensing at linear order \cite{Kamionkowski:2009}. As discussed in \sec{sec:pow}, however, the contributions of the lensing effect appear in the power spectrum of the estimator. 

\subsection{An efficient form of computing the estimator}

Before exploring the reconstruction of the cosmic rotation power spectrum, here we describe an algorithm to efficiently compute the quadratic estimator.  

Let us consider an efficient form of the normalization given in \eq{Eq:norm}. The normalization of \eq{Eq:norm} is expressed as a convolution of two quantities and is rewritten as
\al{
	\frac{1}{A^\a_L} = 2\sum_{i=1}^3 \FT{\hatn}{\bL}{f} 
		\Re[\bm{A}^{E,i} (\hatn) \cdot \bm{A}^{B,i} (\hatn)] 
	\,.
}
where the quantities $\bm{A}^{E,i}(\hatn)$ and $\bm{A}^{B,i}(\hatn)$ are the inverse Fourier transform of the following quantities: 
\al{
	\bm{A}^{E,1}_{\bl} &= \bm{v}_{\bl}\frac{(\tCEE_\l)^2}{\hCEE_\l} \,, & 
	\bm{A}^{B,1}_{\bl} &= \bm{v}^*_{\bl}\frac{1}{\hCBB_\l} \,, \\
	\bm{A}^{E,2}_{\bl} &= \bm{v}_{\bl}\frac{1}{\hCEE_\l} \,, & 
	\bm{A}^{B,2}_{\bl} &= \bm{v}^*_{\bl}\frac{(\tCBB_\l)^2}{\hCBB_\l} \,, \\
	\bm{A}^{E,3}_{\bl} &= -2\bm{v}_{\bl}\frac{\tCEE_\l}{\hCEE_\l} \,, & 
	\bm{A}^{B,3}_{\bl} &= \bm{v}^*_{\bl}\frac{\tCBB_\l}{\hCBB_\l} \,,
}
with $\bm{v}_{\bl}=(1,\E^{4\iu\varphi_{\bl}})$. Compared to directly compute the integral in \eq{Eq:norm}, the above normalization is evaluated more efficiently by employing the fast Fourier transform. 

The unnormalized quadratic estimator of \eq{Eq:uest} is also described as a convolution. Similar to the quadratic estimator of the gravitational lensing potential, an efficient form of \eq{Eq:uest} is given as
\al{
	\uesta_{\bL} &= 2 \sum_{i=1,2} \FT{\hatn}{\bl}{f} \Re[X^{E,i}(\hatn)X^{B,i}(\hatn)]
	\,. \label{Eq:uest:fast}
}
Here, the quantities $X^{E,i}(\hatn)$ and $X^{B,i}(\hatn)$ are the inverse Fourier transform of
\al{
	X^{E,1}_{\bl} &= \tCEE_\l\E^{2\iu\varphi_{\bl}}\frac{\hE_{\bl}}{\hCEE_\l} 
	\,, \\
 	X^{B,2}_{\bl} &= \E^{-2\iu\varphi_{\bl}}\frac{\hB_{\bl}}{\hCBB_\l} 
	\,, \\
	X^{E,2}_{\bl} &= \E^{2\iu\varphi_{\bl}}\frac{\hE_{\bl}}{\hCEE_\l} 
	\,, \\
 	X^{B,2}_{\bl} &= -\tCBB_\l\E^{-2\iu\varphi_{\bl}}\frac{\hB_{\bl}}{\hCBB_\l} 
	\,.
}
In this paper, we apply the above algorithm to reconstruct the cosmic rotation anisotropies.

\section{Understanding reconstructed cosmic-rotation power spectrum} \label{sec:pow}

The reconstructed cosmic-rotation fields, $\esta$, can be used to extract the cosmic rotation spectrum $C_L^{\a\a}$ through its power spectrum, $C_L^{\esta\esta}$. However, the properties of the reconstructed power spectrum is not well studied especially in the case of high-sensitivity polarization experiments. Here we first discuss important contributions to the reconstructed power spectrum and then show impact of these contributions on the reconstructed power spectrum. 

\subsection{Methodology of cosmic-rotation power reconstruction}

Since we work with the quadratic estimator for the cosmic rotation anisotropies, the power spectrum of the reconstructed cosmic rotation is the four-point correlation. The contributions to this four-point correlation can be broken into disconnected and connected (trispectrum) parts as
\al{
	\ave{|\esta_{\bL}|^2} = \ave{|\esta_{\bL}|^2}_{\rm D} + \ave{|\esta_{\bL}|^2}_{\rm C} 
	\,. \label{Eq:4pt}
}

\subsubsection{Disconnected bias}

The most significant contribution comes from the disconnected piece of the four-point correlation (disconnected bias) which consists of the two-point correlations. Similar to the lensing case \cite{Hu:2001kj}, this contribution is given analytically by 
\al{
	\ave{|\esta_{\bL}|^2}_{\rm D} &= A_L^2 \Int{2}{\bl_1}{(2\pi)^2}\Int{2}{\bl_2}{(2\pi)^2} 
		w_{\bL,\bl_1}^\a w_{\bL,\bl_2}^\a
	\notag \\
		& \times (\ave{E_{\bl_1}E_{\bl_2}}\ave{B_{\bL-\bl_1}B_{\bL-\bl_2}}
	\notag \\
		&+ \ave{E_{\bl_1}B_{\bL-\bl_2}}\ave{B_{\bL-\bl_1}E_{\bl_2}})
	\notag \\
		&= \frac{1}{2}\ave{|\esta^{E_{\bm 1},B_{\bm 2}}_{\bL}+\esta^{E_{\bm 2},B_{\bm 1}}_{\bL}|^2}_{\bm 1,2} 
	\,. \label{Eq:4pt:d}
}
Here the index $i={\bm 1}, {\bm 2}$ denotes one of two sets of independent realizations and $\ave{\cdots}_{\bm i}$ is the ensemble average over the $i$th set of realizations. 

In practical analysis, the disconnected bias should be accurately estimated as this term is the most significant source of the four-point correlation. More accurate estimate of the disconnected bias can be realized by replacing part of the simulated $E$ and $B$ modes with the observed values. In this realization-dependent method, the disconnected bias is estimated as
\al{
	(2\pi)^2\delta^D_{\bm{0}} \hN_L^{(0)} 
		&\equiv \ave{|\esta_{\bL}^{E,B_{\bm 1}}+\esta_{\bL}^{E_{\bm 1},B}|^2}_1 
	\notag \\
		&- \frac{1}{2}\ave{|\esta^{E_{\bm 1},B_{\bm 2}}_{\bL}+\esta^{E_{\bm 2},B_{\bm 1}}_{\bL}|^2}_{\bm 1,2} 
	\,. \label{Eq:hN}
}
Here $\delta^D$ is the delta function in the Fourier space. The above estimator is derived as the optimal trispectrum estimator analogues to the lensing case \cite{Namikawa:2012} (see Appendix \ref{app:rdn0} for derivation). Realization-dependent methods are useful to suppress spurious off-diagonal elements in the covariance matrix of the power spectrum estimates and decrease the statistical uncertainties (e.g., \cite{Hanson:2010rp}). In addition, \eq{Eq:hN} is less sensitive to errors in covariance compared to the other approaches \cite{Namikawa:2012}. We use the realization-dependent method to estimate the disconnected bias in the following analysis. 

\subsubsection{N1 term}

Similar to the lensing reconstruction case \cite{Kesden:2003cc}, the trispectrum of \eq{Eq:4pt} is expressed as
\al{
	\ave{|\esta_{\bL}|^2}_{\rm C} = (2\pi)^2\delta^D_{\bm{0}}(C_L^{\a\a}+N_L^{(1)}) + \mC{O}[(C_L^{\a\a})^2] 
	\,. \label{Eq:4pt:c}
}
Additional contributions $N_L^{(1)}$ is usually referred to as the N1 term. The N1 term in the cosmic rotation is given by
\al{
	(2\pi)^2\delta^D_{\bm{0}} N_L^{(1)} &\equiv A_L^2 \Int{2}{\bl_1}{(2\pi)^2}\Int{2}{\bl_2}{(2\pi)^2} 
		w_{\bL,\bl_1}^\a w_{\bL,\bl_2}^\a
	\notag \\
		&\times \ave{\ave{E_{\bl_1}E_{\bl_2}}_\a\ave{B_{\bL-\bl_1}B_{\bL-\bl_2}}_\a
	\notag \\
		&+ \ave{E_{\bl_1}B_{\bL-\bl_2}}_\a\ave{B_{\bL-\bl_1}E_{\bl_2}}_\a}
	\,. 
}
Here $\ave{\cdots}_a$ denotes the ensemble average with a fixed realization of $\a$ and $\ave{\cdots}$ is the usual ensemble average. The N1 term can be efficiently computed by simulation as
\al{
	(2\pi)^2\delta^D_{\bm{0}} N_L^{(1)} 
		&= \frac{1}{2} \ave{|\esta_{\bL}^{E_{\bm 1},B_{\bm 2}}+\esta_{\bL}^{E_{\bm 2},B_{\bm 1}}|^2}_\a
	\notag \\
		&- \frac{1}{2} \ave{|\esta^{E_{\bm 1},B_{\bm 2}}_{\bL}+\esta^{E_{\bm 2},B_{\bm 1}}_{\bL}|^2}_{\bm 1,2} 
	\,.
}
Here the two set of realizations has the same realization of $\a$ but with uncorrelated unrotated CMB. 

\subsubsection{Higher-order term}

The term from higher orders of $C_L^{\a\a}$ could additionally generate the connected part of the four-point correlation \eq{Eq:4pt:c}. The impact of the higher-order terms on the power spectrum estimation would be smaller than that of the N1 term, but the accuracy of the lensing reconstruction from the Planck experiment is already affected by the second order of the lensing potential power spectrum \cite{Hanson:2009dr}. We examine the impact of the higher-order term with the simulation described below. 

\subsubsection{Lensing bias}

The presence of the lensing effect does not biases the estimator of the cosmic rotation anisotropies \cite{Kamionkowski:2009}. However, in the power spectrum analysis, the lensing-induced trispectrum leads to a bias in the power spectrum of the cosmic rotation estimator (hereafter, lensing bias). The lensing signals have been detected by multiple CMB experiments and the impact of the lensing bias should be studied in the cosmic rotation measurement. This bias can be estimated using the standard lensed-$\Lambda$CDM simulations. 

\subsection{Simulation of CMB map}

To explore the impact of the above contributions, we employ the following simulated CMB maps. 

We first compute CMB and lensing-potential angular power spectra using {\tt CAMB} \cite{Lewis:1999bs} and generate unlensed CMB and lensing-potential maps as a random Gaussian fields. We assume $200^\circ\times 200^\circ$ square maps. The CMB maps are then remapped by the lensing potential based on the algorithm of Ref.~\cite{Louis:2013}. 

In addition to the lensed-CMB map, we also create a map including rotation as follows. We generate random fields of anisotropic rotation maps, $\a(\hatn)$, (where $\hatn$ denotes a position) whose power spectrum is described by \eq{Eq:claa:n=0} with $A_{CB}=1$ The CMB polarization maps are then rotated by $\a(\hatn)$ according to \eq{Eq:rotmap}. We denote these maps as ``rotated lensed-CMB'' map.

The instrumental noise is generated as a random Gaussian fields with the following power spectrum \cite{Knox:1999}: 
\al{
	\mC{N}_\l &\equiv \left(\frac{\Delta\rom{P}}{T\rom{CMB}}\right)^2
		\exp\left[\frac{\l(\l+1)\theta^2}{8\ln 2}\right]
	\,. \label{noise}
}
$T\rom{CMB}=2.7$K is the CMB mean temperature. The quantity $\theta$ is beam size and $\Delta\rom{P}$ is the noise level of a polarization map. We consider two cases of the instrumental noise: (1) LiteBIRD (and BK)--like noise, i.e., $\Delta\rom{P}=3\mu$K-arcmin with $\theta=30$ arcmin, and (2) CMBS4-like noise, i.e.,  $\Delta\rom{P}=1\mu$K-arcmin with $\theta=3$ arcmin. 

\subsection{Results of the power spectrum reconstruction}

\subsubsection{Unbiasedness}

\begin{figure}[t]
\bc
\includegraphics[width=8.5cm,height=6cm,clip]{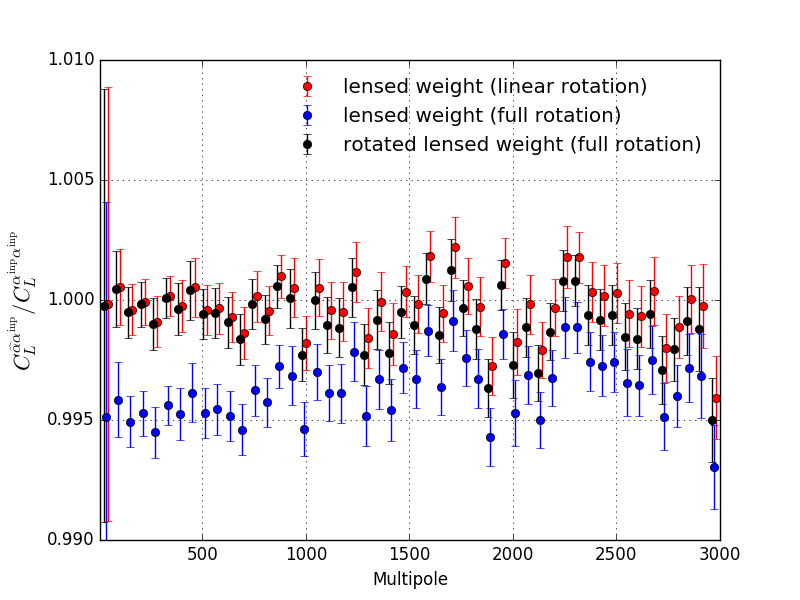} 
\caption{
Cross power spectrum between the input and reconstructed cosmic-rotation fluctuations, divided by the input cosmic-rotation spectrum with the Monte Carlo errors (the standard deviation of $100$ realizations divided by $\sqrt{100}$). The rotated CMB map is created according to \eq{Eq:linrot} (linear) or \eq{Eq:rotmap} (full). The ``lensed weight'' and ``rotated-lensed weight'' denote the cases with the lensed and rotated-lensed CMB spectra in the weight function of \eq{Eq:weight:a}, respectively.
}
\label{Fig:inpxrec}
\ec
\end{figure}

\begin{figure*}[t]
\bc
\includegraphics[width=8.5cm,height=6.5cm,clip]{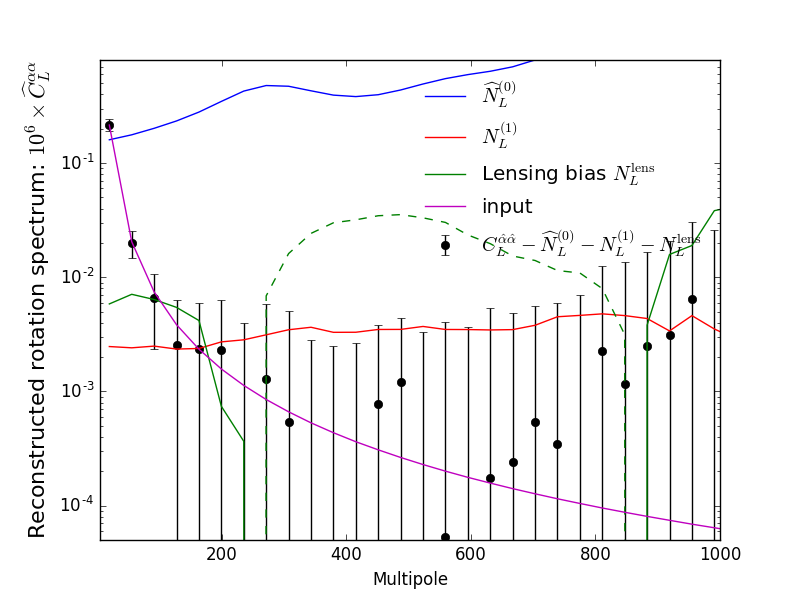} 
\includegraphics[width=8.5cm,height=6.5cm,clip]{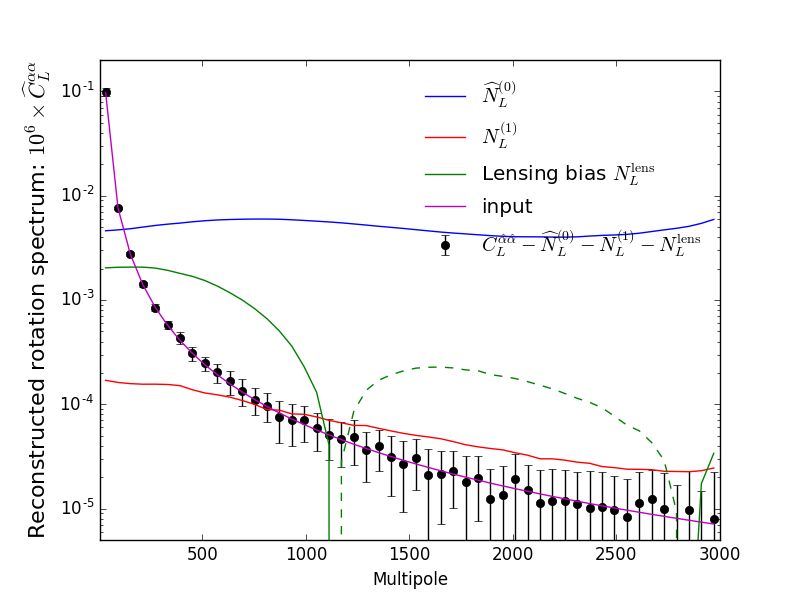} 
\caption{
Angular power spectrum of the reconstructed cosmic rotation with subtraction of the disconnected bias, N1 term and lensing bias (black points) with the $1\sigma$ full-sky measured error in the reconstruction for one realization (Left: BK/LiteBIRD--like noise, Right: CMBS4-like noise). We use $100$ realizations of the simulated maps and the Monte Carlo error is $10\%$ of the error bars. We also show the disconnected bias (blue), N1 term (red), lensing bias (green), and the input rotation power spectrum (magenta). The dashed line shows the negative value. 
}
\label{Fig:caa}
\ec
\end{figure*}

\begin{figure*}[t]
\bc
\includegraphics[width=8.5cm,clip,height=6.5cm]{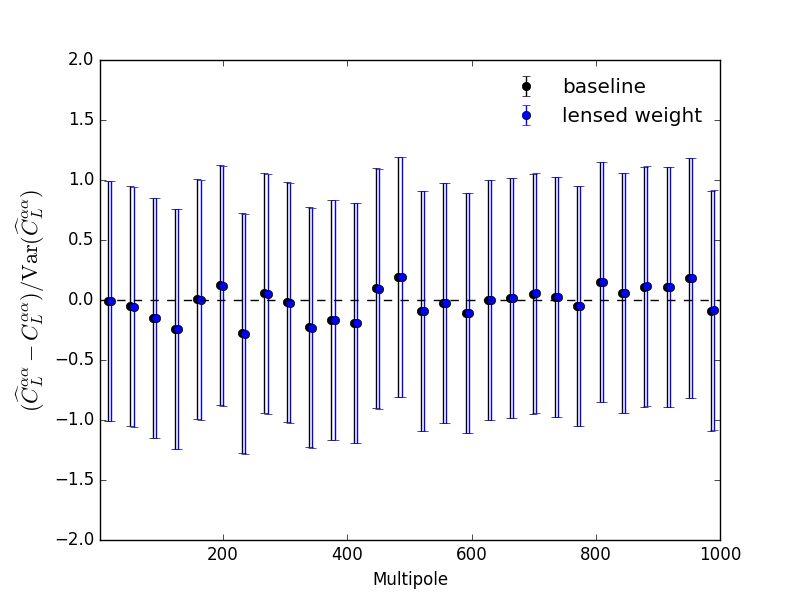} 
\includegraphics[width=8.5cm,clip,height=6.5cm]{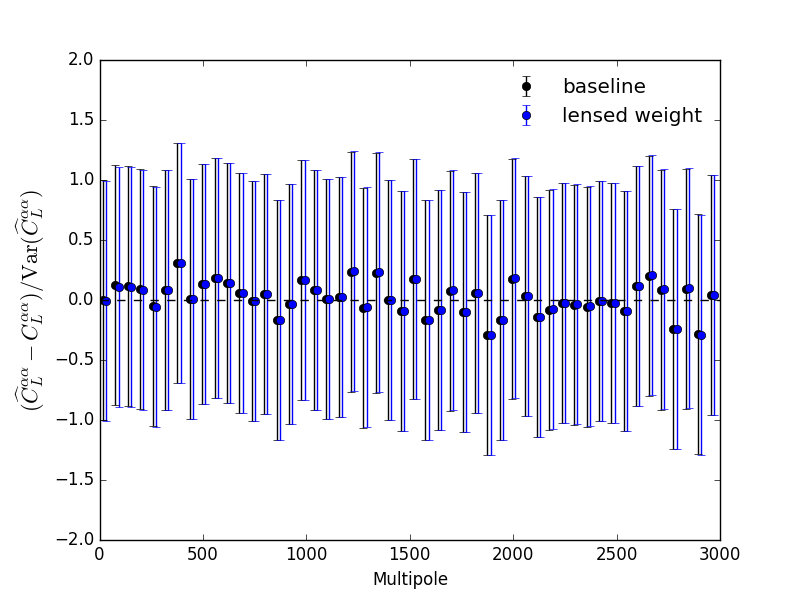} 
\caption{
Angular power spectrum of the reconstructed cosmic rotation after the subtraction of the disconnected bias, N1 term, lensing bias and input cosmic-rotation power spectrum (black) with the $1\sigma$ full-sky measured error for one realization (Left: BK/LiteBIRD--like noise, Right: CMBS4-like noise). We use $100$ realizations of the simulated maps and the Monte Carlo error is $10\%$ of the error bars. We also show the case with the lensed CMB power spectrum in the weight function of \eq{Eq:weight:a} (lensed weight). 
}
\label{Fig:null}
\ec
\end{figure*}

\begin{figure}[t]
\bc
\includegraphics[width=9cm,clip,height=7cm]{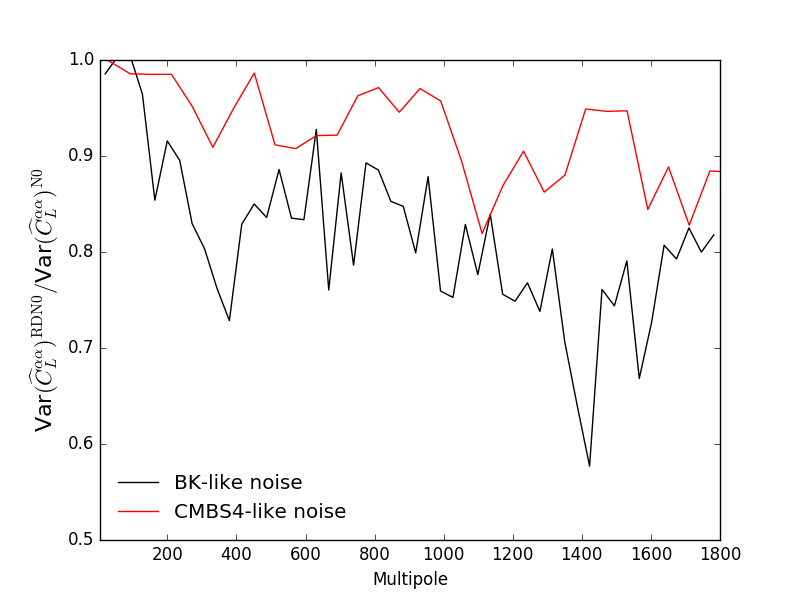} 
\caption{
The variance of the reconstructed rotation spectrum with the realization-dependent disconnected bias (RDN0) divided by that with the analytic disconnected bias (N0). We show the cases with the BK/LiteBIRD--like and CMBS4-like noise. 
}
\label{Fig:rdn0}
\ec
\end{figure}

\begin{figure}[t]
\bc
\includegraphics[width=9cm,clip,height=7cm]{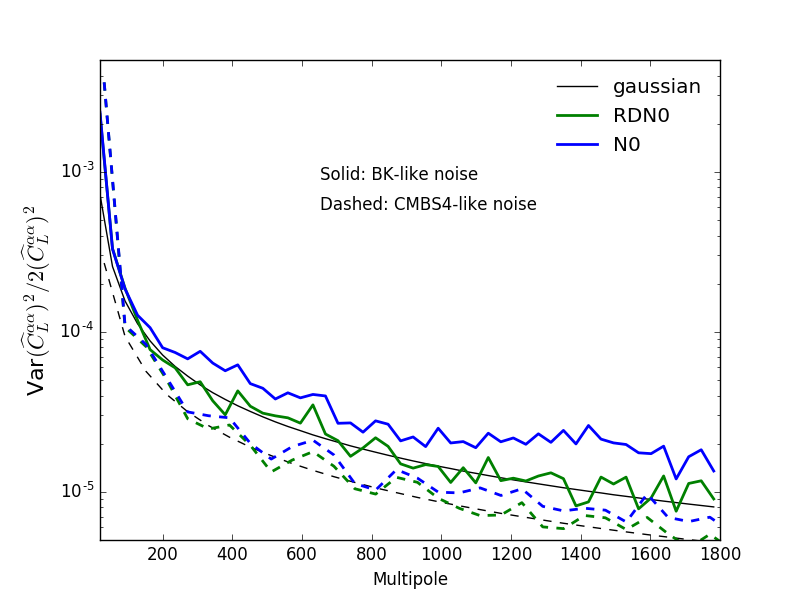} 
\caption{
Test of the Gaussian variance. We show the variances of the reconstructed rotation spectrum with the realization-dependent disconnected bias (RDN0) and with the analytic disconnected bias (N0). They are divided by $2(\hC_L^{\a\a})^2$. The black lines show the case when the reconstructed power spectrum is described as a Gaussian statistics. Both the BK/LiteBIRD--like (solid) and CMBS4-like noise cases (dashed) are plotted. 
}
\label{Fig:var}
\ec
\end{figure}

\begin{figure*}[t]
\bc
\includegraphics[width=8.5cm,clip,height=6.5cm]{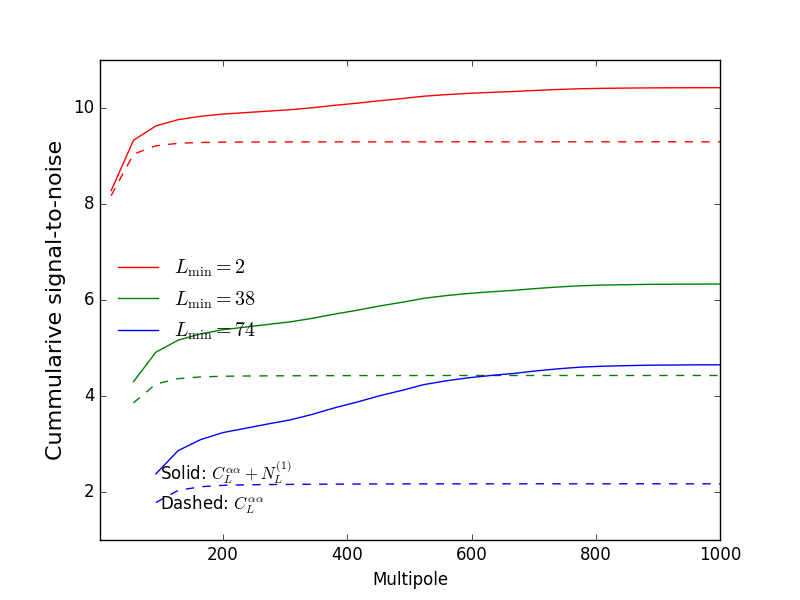} 
\includegraphics[width=8.5cm,clip,height=6.5cm]{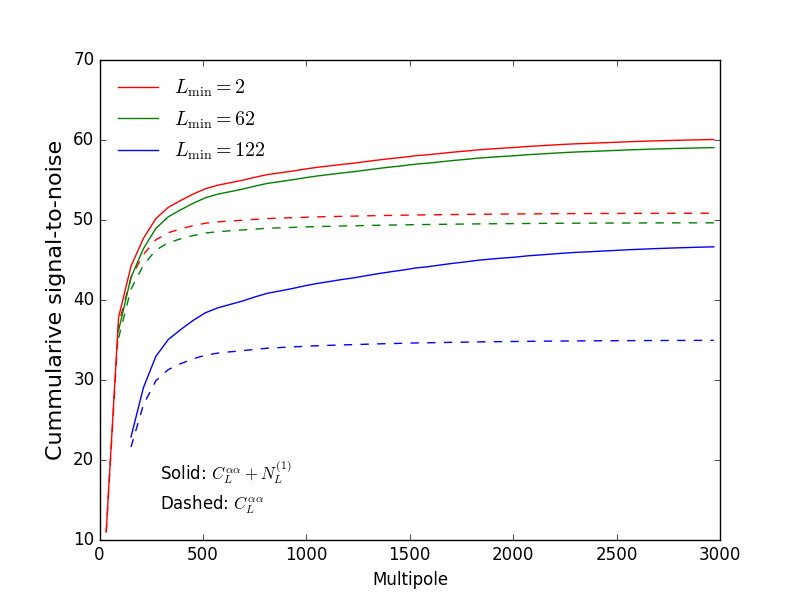} 
\caption{
Cumulative signal-to-noise ratio of the cosmic-rotation power spectrum amplitude with and without the N1 term (Left: BK/LiteBIRD-like noise, Right: CMBS4-like noise). We show the cases with different minimum multipole of the rotation spectrum. 
}
\label{Fig:snr}
\ec
\end{figure*}

We first test the unbiasedness of the cosmic-rotation quadratic estimator given by \eq{Eq:est}. The reconstructed cosmic rotation anisotropies are cross-correlated with the input cosmic rotation fluctuations. This cross-power spectrum is then compared with the input cosmic rotation spectrum. The cross spectrum is given analytically by 
\al{
	(2\pi)^2\delta^D_{\bm{0}} C_L^{\esta\a} &= 
		\Int{2}{\bl}{(2\pi)^2} \frac{w^\a_{\bL,\bl}}{\hCEE_\l\hCBB_{|\bL-\bl|}} 
		\ave{\hE_{\bl}\hB_{\bL-\bl}}_\a \a_{\bL} 
	\notag \\
	&= (2\pi)^2\delta^D_{\bm{0}} C_L^{\a\a} + \mC{O}[(C_L^{\a\a})^2] \,.
}
If the CMB map is rotated by \eq{Eq:linrot}, the above cross spectrum contains up to third order of $\a$ and is equivalent to the input spectrum. 

\fig{Fig:inpxrec} shows the cross power spectrum between reconstructed and input cosmic rotation fluctuations. We show two cases; rotating the polarization map according to \eq{Eq:linrot} (linear) or \eq{Eq:rotmap} (full). We assume the CMBS4-like noise and use $E$ and $B$ modes up to $\l=3000$ in the reconstruction. In the linear case, the cross power spectrum is in good agreement with the input spectrum. If the higher order of $\a$ is included (full), we find that the cross power spectrum is smaller than the input spectrum at sub-percent level. To reduce the higher-order contributions, we follow the similar treatment of the lensing power reconstruction \cite{Lewis:2011fk}; i.e., we use the rotated-lensed power spectrum to the weight function in \eq{Eq:weight:a}. The result is shown as ``rotated-lensed weight''. The corrected cross-power spectrum becomes close to the input spectrum. As we show below, although the impact of the higher-order rotation on the power spectrum reconstruction is not significant, we use the rotated-lensed power spectrum in the baseline analysis. 

\subsubsection{Reconstructed power spectrum}

Next we discuss the cosmic-rotation power spectrum reconstruction employing the simulated CMB maps described above. We perform the power spectrum reconstruction in the two cases of the instrumental noise specification, i.e., the BK/LiteBIRD--like noise and CMBS4-like noise. 

\fig{Fig:caa} shows the power spectrum of the reconstructed cosmic rotation after subtracting the disconnected bias, N1 term and lensing bias, i.e., 
\al{
	\hC_L^{\a\a} = C_L^{\esta\esta} - N_L^{(0)} - N_L^{(1)} - N_L^{\rm lens} \,.
}
Here $N_L^{\rm lens}$ is the lensing bias. This de-biased power spectrum is in good agreement with the input rotation power spectrum in both the BK/LiteBIRD--like and CMBS4-like noise cases. We also show the significance of the disconnected bias (blue), N1 term (red) and lensing bias (green). The most dominant contribution comes from the disconnected bias. The N1 term dominates over the input rotation spectrum at smaller scales ($L=200$ for the BK/LiteBIRD--like noise and $L=800$ for the CMBS4-like noise). Although the lensing bias is smaller than the disconnected bias in both the BK/LiteBIRD--like and CMBS4-like noise cases, the impact of the lensing bias is significant compared to the error bars. Note that the statistical errors are computed for $200\times200$ deg$^2$, and the impact of the lensing bias is reduced for CMB observations at a small patch of sky (e.g., BK). 

\fig{Fig:null} shows difference between the reconstructed power spectrum shown in \fig{Fig:caa} and the input cosmic-rotation power spectrum, $\hC_L^{\a\a}-C_L^{\a\a}$. The difference is consistent with zero within at least sub percent level. This results mean that the reconstructed power spectrum is described by the sum of the disconnected bias, N1 term, lensing bias and the input cosmic-rotation spectrum. We also show that the choice of the weight function does not significantly affect the power spectrum reconstruction, and the reconstructed power spectrum is in good agreement with the sum of the disconnected bias, N1 term, lensing bias and the input cosmic-rotation spectrum. This result implies that, unlike the lensing reconstruction, the higher-order biases such as the ``N2 bias'' \cite{Hanson:2010rp} is negligible in the experimental specifications considered here. 

\subsubsection{Realization-dependent disconnected bias and variance}

The use of the realization-dependent disconnected bias could reduce the statistical uncertainties of the cosmic rotation power spectrum by suppressing the off-diagonal covariance of the reconstructed rotation power spectrum. \fig{Fig:rdn0} shows the variance in the case with the realization-dependent disconnected bias (RDN0) divided by that with the realization-independent disconnected bias (N0). We show the cases with the BK/LiteBIRD--like noise and CMBS4-like noise. In both cases, using the realization-dependent disconnected bias, the statistical uncertainty of the reconstructed power spectrum decreases and is improved by roughly $10$\%--$20$\% compared to the case with the disconnected bias estimated from the simulation alone. 

\fig{Fig:var} shows a test of the Gaussian variance by plotting the following quantity: 
\al{
	N_b \equiv \frac{{\rm Var}(\hC_b^{\a\a})^2}{2(\hC_b^{\a\a})^2} \,. 
}
If the reconstructed power spectrum is Gaussian, $N_b$ coincides with the number of multipoles at the $b$th bin. We find that the realization-dependent bias reduces the non-Gaussian variance of the reconstructed power spectrum. The variance of the reconstructed power spectrum with the realization-dependent bias is consistent with the Gaussian variance. 

\subsubsection{Signal-to-noise}

Since the N1 term is considered as a signal, here we discuss the benefit of adding the N1 term to constrain the comic rotation. We quantify the impact of the inclusion of the N1 term on the cosmic rotation constraints by the following signal-to-noise ratio: 
\al{
	\left(\frac{S}{N}\right)_{<b} 
		= \sqrt{\sum_b\frac{(C_b^{\a\a}+N_b^{(1)})^2}{{\rm Var}(C_b^{\esta\esta})^2} }
	\,. 
}
Here $b$ denotes the bin center of the multipole binning and Var$(C_b^{\esta\esta})$ is the variance of the reconstructed rotation spectrum obtained from simulations. 

\fig{Fig:snr} plots the above signal-to-noise with and without the N1 term. In practice, if CMB data is obtained at some region of sky, CMB analysis is usually performed within the partial sky region and the minimum multipole we can extract is limited. Therefore, we also show the impact of the minimum multipole on the signal-to-noise with varying the minimum multipoles of $C_b^{\a\a}$. Since the cosmic rotation power spectrum becomes large at low multipoles, the signal-to-noise decreases as the minimum multipole becomes large. The impact of the N1 term becomes significant as the minimum multipole increases. The results indicate that, in ongoing and future CMB experiments, to quantify the constraints on the overall amplitude of the cosmic rotation anisotropies, the N1 term is needed to be included. 

\section{Summary} \label{sec:sum}

We have investigated the reconstruction of the cosmic rotation power spectrum from CMB polarization anisotropies, assuming the ongoing and future CMB experiments such as BK, CMBS4, and LiteBIRD. The cosmic rotation power spectrum is assumed to be the scale-invariant spectrum which is motivated by the inflationary origin of the cosmic rotation anisotropies. We found that the N1 term dominates over the original input rotation spectrum at small scales and increases the total signal-to-noise of the amplitudes of the cosmic rotation power spectrum. The lensing bias is significant compared to the statistical error in the case of LiteBIRD and CMBS4, but the impact of the lensing bias becomes negligible for CMB observations at a small patch of sky. The higher-order biases beyond the N1 term is found to be not significant. We showed that the sum of the disconnected bias, N1 term, lensing bias and input rotation spectrum is in good agreement with the power spectrum of the quadratic estimator. We also found that the use of the realization-dependent disconnected bias decreases the statistical uncertainties of the reconstructed rotation power spectrum by $10$\%--$20$\% depending on experimental specifications. 

For high-sensitivity CMB experiments, the lensing $B$ mode degrades the sensitivity to the cosmic rotation \cite{Yadav:2009}. Since such high-sensitivity experiments can significantly suppress the contributions of the lensing $B$ mode by the so-called delensing technique (e.g., \cite{Kesden:2002ku}). However, the $B$ mode delensing at small scales suffers from the delensing bias \cite{Seljak:2003pn}, and demonstration of the cosmic rotation reconstruction from the delensed $B$ modes is required. We leave the analysis including the delensing to future work. 

\begin{acknowledgments}
T.N. is grateful to Chao-Lin Kuo for helpful discussions and support of this work, and also to Vera Gluscevic, Mark Kamionkowski and Brian Keating. This research used resources of the National Energy Research Scientific Computing Center, which is supported by the Office of Science of the U.S. Department of Energy under Contract No. DE-AC02-05CH11231.

\end{acknowledgments}

\onecolumngrid
\appendix

\section{Disconnected bias estimation} \label{app:rdn0}

Here we briefly summarize the derivation of the realization-dependent disconnected bias in the case of the cosmic rotation as described in \eq{Eq:hN}. The derivation is similar to that in the case of the lensing reconstruction which is given in, e.g., the Appendix of Ref.~\cite{BKVIII}. 

The disconnected bias of \eq{Eq:hN} emerges naturally when deriving the optimal trispectrum estimator from the CMB polarization likelihood. The optimal trispectrum estimator is derived from the Edgeworth expansion of the CMB polarization likelihood; 
\al{
	\mC{L} \propto \left[\prod_{i=1}^4\Int{2}{\bl_i}{(2\pi)^2}\right]
		\ave{E_{\bl_1}B_{\bl_2}E_{\bl_3}B_{\bl_4}}_{\rm C}
		\PD{}{E_{\bl_1}}\PD{}{B_{\bl_2}}\PD{}{E_{\bl_3}}\PD{}{B_{\bl_4}} \mC{L}_{\rm g}
	\,. 
}
Here $\mC{L}_{\rm g}$ is the Gaussian likelihood of the $E$ and $B$ mode. The trispectrum induced by the cosmic rotation is given as 
\al{
	\ave{E_{\bl_1}B_{\bl_2}E_{\bl_3}B_{\bl_4}}_{\rm C}
	\propto \delta^D_{\bl_1+\bl_2+\bl_3+\bl_4}
		[w^\a_{\bl_1+\bl_2,\bl_1} w^\a_{\bl_3+\bl_4,\bl_3} C_{|\bl_1+\bl_2|}^{\a\a}
		+ w^\a_{\bl_1+\bl_4,\bl_1} w^\a_{\bl_2+\bl_3,\bl_2} C_{|\bl_1+\bl_4|}^{\a\a}]
	\,,
}
where $w^\a_{\bL,\bl}$ is defined in \eq{Eq:weight:a}. The approximate formula of the estimator which is numerically tractable is proportional to the derivative of the log-likelihood with respect to $C_L^{\a\a}$. The derivative is given by
\al{
	\PD{\mC{L}}{C^{\a\a}_L} 
	\propto \left[\prod_{i=1}^2\Int{2}{\bl_i}{(2\pi)^2}\right] w^\a_{\bL,\bl_1} w^\a_{-\bL,\bl_2}
		\PD{}{E_{\bl_1}}\PD{}{B_{\bL-\bl_1}} \PD{}{E_{\bl_2}}\PD{}{B_{-\bL-\bl_2}} \mC{L}_{\rm g}
	\simeq \left(|\uesta^{EB}_{\bL}|^2 - \frac{\hN^\a_{\bL}}{(A^\a_{\bL})^2} \right) \mC{L}_{\rm g}
	\,. 
}
After correcting the normalization, the above equations leads to an approximate formula of the optimal estimator for $C_L^{\a\a}$ with the subtraction of the disconnected bias. 

\bibliographystyle{apsrev}
\bibliography{main,exp}

\end{document}